\documentclass[aps,pra,twocolumn,a4paper,pra,superscriptaddress]{revtex4}


\usepackage{amsmath}
\usepackage{amssymb}
\usepackage{bm}

\usepackage{color,epsfig,graphicx}

\newcommand{\ket}[1]{\left|#1\right>}
\newcommand{\bra}[1]{\left<#1\right|}
\newcommand{\braket}[2] {\left< \left. #1\vphantom{#2} \right| #2 \right>} 
\newcommand{\Set}[1]{\left\lbrace #1 \right\rbrace}
\def\cH{ {\ensuremath{\mathcal{H}}} }
\def\cHC{ \ensuremath{\cH^C} }
\def\cHG{ \ensuremath{\cH^G} }
\def\Id{\openone}
\def\cO{ {\ensuremath{\mathcal{O}}} }
\def\tdelta[#1]{\ensuremath{\tilde{\delta}_{#1}}}

\def\etamax{\eta_{\mathrm{max}}}
\def\mD{ {\mathcal D} }
\def\vrho{ \ket{0}\!\bra{0} }
\def\xtg{\ensuremath{x_{\mathrm{t}}}}
\def\txtg{\ensuremath{{\tilde x}_{\mathrm{t}}}}
\newcommand{\Tr}{\mathop{\mathrm{Tr}}\nolimits}

\newcommand{\acot}{\mathop{\mathrm{acot}}\nolimits}

\begin{document}

\title{Scattering quantum random-walk search with errors}

\author{A.~G\'abris}
\affiliation{
Research Institute for Solid State Physics and Optics, H-1525
  Budapest, P. O. Box 49, Hungary
}
\author{T.~Kiss}
\affiliation{
Research Institute for Solid State Physics and Optics, H-1525
  Budapest, P. O. Box 49, Hungary
}
\author{I.~Jex}
\affiliation{
Department of Physics, FJFI {\v C}VUT, {B\v r}ehov\'a 7, 115
  19 Praha 1 - Star{\'e} M\v{e}sto, Czech Republic
}
\date{\today}
\begin{abstract}
  We analyze the realization of a quantum-walk search algorithm in a
  passive, linear optical network. The specific model enables us to
  consider the effect of realistic sources of noise and losses on the
  search efficiency. Photon loss uniform in all directions is shown to
  lead to the rescaling of search time.  Deviation from directional
  uniformity leads to the enhancement of the search efficiency
  compared to uniform loss with the same average. In certain cases
  even increasing loss in some of the directions can improve search
  efficiency. We show that while we approach the classical limit of
  the general search algorithm by introducing random phase
  fluctuations, its utility for searching is lost. Using numerical
  methods, we found that for static phase errors the averaged search
  efficiency displays a damped oscillatory behaviour that
  asymptotically tends to a non-zero value.
\end{abstract}

\maketitle

\section{Introduction}

The generalization of random walks for quantum systems
\cite{AHARONOV1993QRW} proved to be a fruitful concept
\cite{kempe-2003-44} attracting much recent interest. Algorithmic
application for quantum information processing is an especially
promising area of utilization of quantum random walks (QRW)
\cite{Ambainis2004Quantum-walks-a}.

In his pioneering paper \cite{Grover} Grover presented a quantum
algorithm that can be used to search an unsorted database
quadratically faster than the existing classical algorithms. Shenvi,
Kempe and Whaley (SKW) \cite{shenvi:052307} proposed a search
algorithm based on quantum random walk on a hypercube, which has
similar scaling properties as the Grover search. In the SKW algorithm
the oracle is used to modify the quantum coin at the marked vertex. In
contrast to the Grover search, this algorithm generally has to be
repeated several times to produce a result, but this merely adds a
fixed overhead independent of the size of the search space.

There are various suggestions and some experiments how to realize
quantum walks in a laboratory. The schemes proposed specifically for
the implementation of QRWs include ion traps
\cite{travaglione:032310}, nuclear magnetic resonance
\cite{du-2003-67} (also experimentally verified
\cite{Ryan2005Experimental-im}), cavity quantum electrodynamics
\cite{di:032304,agarwal:033815}, optical lattices \cite{dur-2002-66},
optical traps \cite{eckert:012327}, optical cavity
\cite{Roldan2005Optical-impleme}, and classical optics
\cite{knight-2003-227}. Moreover, the application of standard general
logic networks to the task is always at hand \cite{Hines-pra75,
Fujiwara2005Scalable-networ}.

The idea of the scattering quantum random walk (SQRW)
\cite{hillery:032314} was proposed as an answer to the question that
can be posed as: how to realize a coined walk by a quantum optical
network built from passive, linear optical elements such as beam
splitters and phase shifters? It turned out that such a realization is
possible and, in fact, it leads to a natural generalization of the
coined walk, the scattering quantum random walk
\cite{Kosik2005Scattering-mode}. The SQRW on the hypercube allows for
a quantum optical implementation of the SKW search algorithm
\cite{shenvi:052307}. Having a proposal for a physical realization at
hand we are in the position to analyze in some detail the effects
hindering its successful operation.

Noise and decoherence strongly influence quantum walks. For a recent
review on this topic see \cite{Kendon2006Decoherence-in-}. The first
investigations in this direction indicated that a small amount of
decoherence can actually enhance the mixing property
\cite{Kendon2003Decoherence-can}.  For a continuous QRW on a hypercube
there is a threshold for decoherence, beyond which the walk behaves
classically \cite{Alagic2005Decoherence-in-}. Ko\v sik {\it et al}
analyzed SQRW with randomized phase noise on a $d$ dimensional lattice
\cite{Kosik2006Quantum-walks-w}. The quantum walk on the line has been
studied by several authors in the linear optical context, with the
emphasis on the effect of various initial states, as well as on the
impact of decoherence \cite{jeong:pra68.012310,
pathak:pra75.032351}. The quantum random walk search with imperfect
gates was discussed in some detail by Li {\it et al}
\cite{Li2006Gate-imperfecti}, who have considered the case when the
Grover operator applied in the search is systematically modified. Such
an imperfection decreases the search probability and also shifts its
first maximum in time.

In this paper we analyze the impact of noise on the SKW algorithm
typical for the experimental situations of the SQRW. In particular,
first we focus on photon losses and show that, somewhat contradicting
the na\"{\i}ve expectation, non-trivial effects such as the
enhancement of the search efficiency can be observed.  As a second
type of errors we study randomly distributed phase errors in two
complementary regimes.  The first regime is characterized by rapid
fluctuation of the optical path lengths, that leads to the
randomization of phases for each run of the algorithm. We show that
the classical limit of the SKW algorithm, reached by increasing the
variance of the phase fluctuations, does not correspond to a search
algorithm. In the other regime, the stability of the optical path
lengths is maintained over the duration of one run, thus the errors
are caused by static random phases. This latter case has not yet been
considered in the context of QRWs. We found that static phase errors
bring a significantly different behaviour compared to the case of
phase fluctuations. Under static phase errors the algorithm retains
its utility, with the average success probability displaying a damped
oscillatory behaviour that asymptotically tends to a non-zero constant
value.

The paper is organized as follows. In the next section we introduce
the scattering quantum walk search algorithm. In section
\ref{sec:uniform}.\ we derive analytic results for the success
probability of search for the case when a single coefficient describes
photon losses independent of the direction. In section
\ref{sec:non-uniform}.\ we turn to direction dependent losses, and
present estimations of the success probability based on analytical
calculations and numerical evidence. In section \ref{sec:phase}.\
phase noise is considered and consequences for the success probability
are worked out. Finally, we conclude in Sec.~\ref{sec:conclusions}.

\section{The scattering quantum walk search algorithm}
\label{sec:sqrw}

The quantum walk search algorithm is based on the generalized notion
of coined quantum random walk (CQRW), allowing the coin operator to
be non-uniform accross the vertices. In the early literature the coin
is considered as position (vertex) independent. The CQRW is defined on
the product Hilbert space $\cH=\cHC \otimes \cHG$, where $\cHC$ refers
to the quantum coin, and $\cHG$ represents the graph on which the
walker moves. The discrete time-evolution of the system is governed by
the unitary operator
\begin{equation}
U=SC\, ,
\end{equation}
where $C$ is the coin operator which corresponds to flipping the
quantum coin, and $S$ is the step or translation operator that moves
the walker one step along some outgoing edge, depending on the coin
state. Adopting a binary string representation of the vertices $V$ of
the underlying graph $G=(V,E)$, the step operator $S$ (a permutation
operator of the entire Hilbert space $\cH$) can be expressed as
\begin{equation}
S = \sum_{d=0}^{n-1} \sum_{x\in V} \ket{d, x\oplus e_{dx}} \bra{d,x}\, .
\label{eq:S_gendef}
\end{equation}
In (\ref{eq:S_gendef}) $x$ denotes the vertex index. Here, and in the
rest of this paper we identify the vertices with their indices and
understand $V$ as the set of vertex indices. The most remarkable fact
about $S$ is that it contains all information about the topology of
the graph. In particular, the actual binary string values of $e_{dx}$
are determined by the set of edges $E$. This is accomplished by the
introduction of direction indices $d$, which run from $0$ to $n-1$ in
case of the $n$ regular graphs which are used in the search algorithm.

To implement the scattering quantum random walk on an $n$ regular
graph of $N$ nodes, identical $n$-multiports \cite{Jex-optcomm117,
Zukovski_pra55} are arranged in columns each containing $N$
multiports. The columns are enumerated from left to right, and each
row is assigned a number sequentially. The initial state enters on the
input ports of multiports in the leftmost column. The output and input
ports of multiports of neighbouring columns $j$ and $j+$ are then
indexed suitably and connected according to the graph $G$.

For the formal description of quantum walks on arrays of multiports,
we propose to label every mode by the row index and input port index
of its \textit{destination} multiport. We note that an equally good
labelling can be defined using the row index and output port index of
the \textit{source} multiport. To describe single excitation states,
we use the notation $\ket{d,x}$ where the input port index of the
destination multiport is $d=0,1,\ldots,n-1$, and the row index is
$x=0,1,\ldots,N-1$. Thus the total Hilbert space can effectively be
separated into some product space $\cHC \otimes \cHG$.  To be precise,
the additional label $j$ would be necessary to identify in which
column the multiport is, however, we think of the column index as a
discrete time index, and drop it as an explicit label of modes. Thus a
time-evolution $U=SC$ can be generated by the propagation through
columns of multiports.

\begin{figure}
\includegraphics[scale=.85]{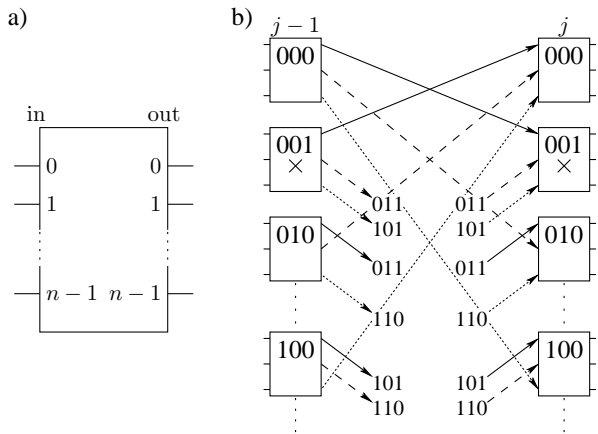}
\caption{a) Illustration of the labelling of input and output ports of
multiports used for the realization of the walk on the $n$ dimensional
hypercube. b) Schematic depiction of the setup of the SQRW
implementation of the SKW algorithm for $n=3$, with the marked node
being $\xtg=001$. For clarity, a unique line pattern is associated
with each input (output) port index.  }
\label{fig:port-example}
\end{figure}

A quantum walk can be realized in terms of the basis defined using the
destination indices, and we shall term it ``standard basis'' through
this section. First, we recall that an $n$-multiport can be fully
characterized by an SU($n$) transformation matrix $\mathbf{C}$.  The
effect of such multiport on single excitation states
$\ket{\psi}\in\cHC$ is given by the formula,
\begin{equation}
  \ket{\psi}=\sum_{d=0}^{n-1} a_d \ket{d} \to \sum_{d,k=0}^{n-1}
  C_{dk} a_k \ket{d}, 
\label{eq:C}
\end{equation}
where $\ket{d}$ denotes the single photon state with the photon being
in the $d$ mode, i.e.\ $\ket{d} = \ket0_0 \ldots \ket1_d \ldots
\ket0_{n-1}$. We note, that a multiport with any particular
transformation matrix $\mathbf{C}$ can be realized in a laboratory
\cite{Reck_prl73}. To simplify calculations it may be beneficial to
choose an indexing of input and output ports such that the connections
required to realize the graph $G$ can be made in such way that each
input port has the same index as the corresponding source output
port. Therefore the label $d$ can stay unique during ``propagation.''
We emphasize that this is not a necessary assumption for a proper
definition of SQRW, but an important property that makes also easier
to see that SQRWs are a superset of generalized CQRWs. This indexing
of input and output ports for walks on a hypercube is depicted on
Fig.~\ref{fig:port-example}a, with some of the actual connections
illustrated for a (three dimensional) cube on
Fig.~\ref{fig:port-example}b.

Considering an array of identical multiports, an arbitrary input state
undergoes the transformation by the same matrix $\mathbf{C}$ for every
$x$. Let the output port $d$ of multiport $x$ be connected to
multiport $x\oplus e_{dx}$ in the next row. Thus the mode labelled by
the source indices $d$ and $x$, is labelled by $d$ and $x\oplus
e_{dx}$ in terms of the destination indices. Therefore, effect of
propagation in terms of our standard basis is written,
\begin{equation}
\sum_{d,x} a_{dx} \ket{d,x} \to \sum_{dkx} C_{dk} a_{kx} \ket{d,x
  \oplus e_{dx}}.
\label{eq:SC}
\end{equation} 
Comparing this formula with Eqs.~(\ref{eq:S_gendef}) and (\ref{eq:C})
we see that this formula corresponds to a $U=SC=S(C_0\otimes\Id)$
transformation where $C_0$ is generated by the matrix $\mathbf{C}$.
Due to the local nature of the realization of the coin operation, it
is straight-forward to realize position dependent coin operations,
such as the one required for the quantum walk search algorithm.

In particular, the SKW algorithm \cite{shenvi:052307} is based on the
application of two distinct coin operators, e.g.
\begin{subequations}
\label{eq:std-coin-pair}
\begin{eqnarray}
C_0 &=& G, \\
C_1 &=& -\Id,
\end{eqnarray}
\end{subequations}
where $G$ is the Grover inversion or diffusion operator $G:= -\Id + 2
|s^C\rangle\langle s^C|$, with $|s^C\rangle=1/\sqrt{n} \sum_{d=1}^{n}
\ket{d}$ \cite{moore02quantum}. In the algorithm, the application of
the two coin operators is conditioned on the result of oracle operator
$\cO$. The oracle marks one $\xtg$ as target, hence the coin operator
becomes conditioned on the node:
\begin{equation}
C' = C_0 \otimes \Id + (C_1 - C_0)\otimes \ket{\xtg}\bra{\xtg}.
\label{eq:pert_coin}
\end{equation}
When $n$ is large, the operator $U':=SC'$ can be regarded as a
perturbed variation of $U=S(C_0\otimes\Id)$. The conditional
transformation (\ref{eq:pert_coin}) is straight-forward to implement
in the multiport network. For the two coins (\ref{eq:std-coin-pair})
one has to use a simple phase shifter at position $\xtg$ in every
column of the array, and a multiport realizing the Grover matrix $G$
at every other position. The connection topology required to implement
a walk on the hypercube is such that in the binary representation we
have $e_d=0\ldots1\ldots0$ with 1 being at the $d$'th position, i.e.\
$e_d=2^d$. See Fig.~\ref{fig:port-example}b for a schematic example,
when $\xtg=001$.

The above described scheme to realize quantum walks in an array of
multiports using as many columns as the number of iterations of $U$
can be reduced to only a single column. To do this, one simply needs
to connect the output ports back to the appropriate input ports of the
destination multiport in the same column. This feed-back setup is
similar to the one introduced in Ref.~\cite{Kosik2005Scattering-mode}.

\section{Uniform decay}
\label{sec:uniform}

\begin{figure}
\includegraphics[scale=.67]{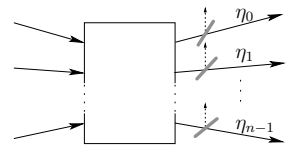}
\caption{Schematic illustration of the photon loss model being
  used. The losses suffered by each output mode are represented by
  fictitious beam-splitters with transmittivities $\eta_d$. The
  beam-splitters incorporate the combined effect of imperfections of
  the multiport devices, and effects influencing the state during
  propagation between the multiports (e.g.\ scattering and
  absorption).}
\label{fig:loss-scheme}
\end{figure}

We begin our analysis of the effect of errors on the quantum walk
search algorithm by concentrating on photon losses. In an optical
network, photon losses are usually present due to imperfect optical
elements. An efficient model for linear loss is to introduce
fictitious beam-splitters with transmittances corresponding to the
effective transmission rate (see Fig.~\ref{fig:loss-scheme}).

The simplest case is when all arms of the multiports are characterized
by the same linear loss rate $\eta$. The operator describing the
effect of decay on a single excitation density operator can then be
expressed as
\begin{equation}
\mD (\varrho) = \eta^2 \varrho + (1-\eta^2)\vrho.
\label{eq:homloss_op}
\end{equation}
The total evolution of the system after one iteration may be written
as $\varrho \to \mD(U\varrho U^{\dag})$. It is important to note that
with the introduction of this error, the original Hilbert space $\cHG$
of one-photon excitations must be extended by the addition of the
vacuum state $\ket0$. The action of the SQRW evolution operator $U$ on
the extended Hilbert space follows from the property
$U\ket0=\ket0$. Due to the nature of Eq.~(\ref{eq:homloss_op}) and the
extension of $U$, one can see that the order of applying the unitary
time step and the error operator $\mD$ can be interchanged. Therefore,
over $t$ steps the state of the system undergoes the transformation
\begin{equation}
\varrho \to \eta^{2t} U^t \varrho U^{\dag t}+ (1-\eta^{2t})\vrho = \mD^t(U^t
\varrho U^{\dag t}). 
\end{equation}

To simplify calculations, we introduce a linear (but non-unitary)
operator to denote the effect of the noise operator $\mD$ on the
search Hilbert space:
\begin{equation}
D \ket{\psi} = \eta \ket{\psi}.
\end{equation}
This operator is simply a multiplication with a number. It is
obviously linear, however, for $\eta<1$ not unitary. The operator $D$
does not describe any coherence damping within the one-photon
subspace, since it only uniformly decreases the amplitude of the
computational states and introduces the vacuum. Since all final
statistics are gathered from the search Hilbert space $\cHG$, it is
possible to drop the vacuum from all calculations, and incorporate all
information related to it into the norm of the remaining state. In
other words, we can think of $DU$ as the time step operator, and relax
the requirement of normalization. Using this notation, the effect of
$t$ steps is very straight-forward to express:
\begin{equation}
\ket{\psi} \to \eta^t U^{t} \ket{\psi}.
\label{eq:homo_nstep}
\end{equation}
This formula indicates that inclusion of the effect of uniform loss
may be postponed until just before the final measurement. The losses,
therefore, may simply be included in the detector efficiency (using an
exponential function of the number of iterations).

Applying the above model of decay to the quantum walk search algorithm
we define the new step operator $U''=DU'$, and write the final state
of the system after $t$ steps as
\begin{multline}
  (U'')^t \ket{\psi_0} = \\
\eta^t \cos(\omega'_0t) \ket{\psi_0}- \eta^t
  \sin(\omega'_0t)\ket{\psi_1} 
  + \eta^t O\left(\frac{n^{3/4}}{\sqrt{2^n}}\right)
  \ket{\tilde{r}}. 
\end{multline}
Adopting the notation of Ref.~\cite{shenvi:052307}, the probability of
measuring the target state $\ket{x=0}$ at the output after $t$ steps
can be expressed as
\begin{multline}
p_n(\eta,t) = \sum_{d=0}^{n-1} \left|\left< d,0 \left| (U'')^t \psi_0
    \right.\right> \right|^2 \\ 
= \eta^{2t} \sin^2 (\omega'_0 t)
    \left|\left<\left. R,0 \right| \psi_1\right> \right|^2 + 2^{-n}
    \eta^{2t} \cos^2 (\omega'_0t) \\
+ O(1/2^n).
\label{eq:prob_uniform}
\end{multline}
We know from Ref.~\cite{shenvi:052307} that $\left|\left<\left.
R,0\right| \psi_1 \right>\right|^2 = 1/2 - O(1/n)$. Since an overall
exponential drop of the success probability is expected due to the
$\eta^{2t}$ factor, we search for the maximum $t_f$ be before the
ideal time-point $|\omega'_0|t=\pi/2$. This guarantees that $\sin^2
(\omega'_0 t_f)$ is finite, therefore due to the $2^{-n}$ factor for
large $n$ the second term can be omitted, and it is sufficient to
maximize the function
\begin{equation}
p_n(\eta,t) = \eta^{2t} \sin^2 (\omega'_0t) \left( 1/2 - O(1/n) \right),
\label{eq:prob_uniform_approx}
\end{equation}
with respect to $t$. After substituting the result $|\omega'_0| =
1/\sqrt{2^{n-1}}[ 1 - O(1/n) ] \pm O(n^{3/2}/2^n)$ from
Ref.~\cite{shenvi:052307}, these considerations yield the global
maximum at $t_f = \sqrt{2^{n-1}} \left[\acot (-\ln\eta\sqrt{2^{n-1}})
+ O(1/n) \right]$. During operation we set
\begin{equation}
t_m:= \sqrt{2^{n-1}} \acot(-\ln\eta\sqrt{2^{n-1}}), 
\end{equation}
or the closest integer, as the time yielding the maximum probability
of success.

\begin{figure}

\includegraphics[scale=.67]{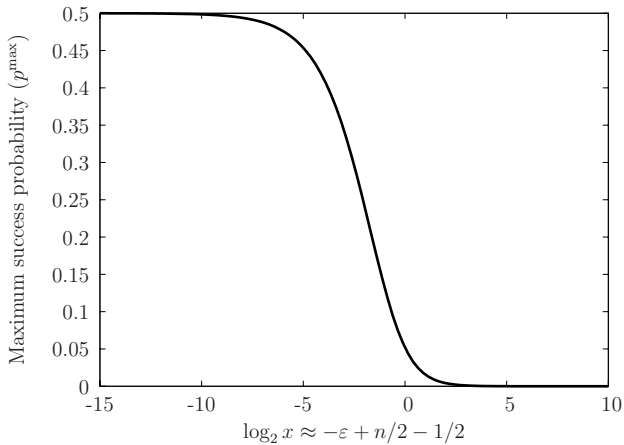}
\caption{Probability of measuring the target state after the optimal
  number of iterations according to the approximation in
  Eq.~(\ref{eq:maxprob-approx}). The probability is plotted against
  the logarithm of $x$ which is a combination of the rank of the
  hypercube $n$ and the logarithmic transmission parameter
  $\varepsilon$.}
\label{fig:plot_prob_dimless}
\end{figure}

\begin{figure}

\includegraphics[scale=.67]{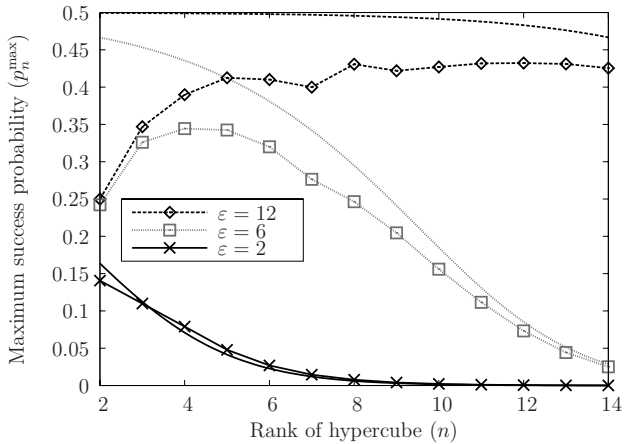}
\caption{Maximum success probabilities for different ranks of
  hypercube ($n$) calculated using the theoretical approximation, and
  numerical simulations. The theoretical curves are drawn with
  continuous lines of different patterns, and the numerical results
  are represented by points interconnected with the same line pattern
  and colour as the theoretical approximates corresponding to the same
  logarithmic transmission parameter $\varepsilon$.}
\label{fig:plot_max_prob_compare}
\end{figure}

To simplify the upcoming formulae, we introduce the variables
\begin{eqnarray}
x &=& -\ln\eta\ \sqrt{2^{n-1}}, \\
\varepsilon &=& \log_2(1 - \eta).
\end{eqnarray}
The variable $\varepsilon$ can be regarded as a logarithmic
transmission parameter (the ideal case corresponds to
$\varepsilon=\infty$, and complete loss to $\varepsilon=0$). When
$\varepsilon$ is sufficiently large, the expression $-\ln\eta$ can be
approximated to first order in $2^{-\varepsilon}$ and we obtain
\begin{equation}
x \approx 2^{-\varepsilon + n/2 - 1/2}.
\label{eq:x-approx}
\end{equation}
Upon substituting $t_m$ into (\ref{eq:prob_uniform_approx}) we can use
the new variable $x$ to express the sine term as
\begin{multline}
  \sin^2(\omega'_0t_m) =\sin^2(|\omega'_0|t_m) =
  \sin^2\left[\acot x(1+O(1/n))\right] =\\
  = \frac1{1+x^2} + \frac{2x\acot
    x}{1+x^2}O(1/n) + \frac{\acot^2x}{1+x^2} O(1/n^2).
\end{multline}
Thus for the maximum success probability $p_n^{\mathrm{max}}(\eta) =
p_n(\eta, t_m)$ we obtain
\begin{equation}
p_n^{\mathrm{max}}(\eta) = \frac{e^{-2x\acot x}}{ 1 + x^2} \left[
  \frac12 - O(1/n) + x\acot x\, O(1/n) \right].
\label{eq:maxprob-homo}
\end{equation}
This formula is our main result for the case of uniform photon
losses. In the large $n$ limit it gives the approximate performance of
the SKW search algorithm as a function of the transmission rate and
the size of the search space. Since $x\acot x$ is bounded in $x$, the
accuracy of the term in brackets is bounded by $O(1/n)$. The most
notable consequence of the second $O(1/n)$ contribution is that while
in the ideal case the probability $1/2$ is an upper bound, in the
lossy case deviations from the leading term,
\begin{equation}
p^{\mathrm{max}}(x) = \frac12 \exp(-2x\acot x) \frac1{1+x^2},
\label{eq:maxprob-approx}
\end{equation}
can be expected in either direction. The functional form of
Eq.~(\ref{eq:maxprob-approx}), plotted on
Fig.~\ref{fig:plot_prob_dimless}, allows for a universal
interpretation of the dependence of success probability on the
transmission rate and the size of the search space through the
combined variable $x$. For small losses we can use the approximation
(\ref{eq:x-approx}) and conclude that the search efficiency depends
only on the difference $n/2-\varepsilon$. The approximation is
compared with the results of numerical calculations on
Fig.~\ref{fig:plot_max_prob_compare}. We can observe the $O(1/n)$
accuracy of the theoretical curves as expected, hence producing poorer
fits at smaller ranks. The positive deviations from the theoretical
curves observable at low transmission rates are due to the second
$O(1/n)$ term of Eq.~(\ref{eq:maxprob-homo}).

\section{Direction dependent loss}
\label{sec:non-uniform}

In the present section we no longer assume equal loss rates, and
consider the schematically depicted loss model on
Fig.~\ref{fig:loss-scheme} with arbitrary $\eta_d$ parameters. Because
of the high symmetry of the hypercube graph, and the use of mainly
identical multiports, we can neglect the position dependence of the
transmission coefficients. The operator $\mD$ describing the
decoherence mechanism thus acts on a general term of the density
operator as
\begin{equation}
\mD(\ket{d,x}\bra{d',x'}) = \eta_d \eta_{d'} \ket{d,x}\bra{d',x'} +
\delta_{xx'} \delta_{dd'} \eta_d^2 \vrho.
\end{equation}
To describe the overall effect of this operator on a pure state, we
re-introduce the linear decoherence operator in a more general form,
\begin{equation}
D = \sum_{d} \eta_d \ket{d}\bra{d} \otimes \Id,
\label{eq:D-inhomo}
\end{equation}
and use the notation $\{\eta\}$ to denote the set of coefficients
$\eta_d$. Due to the symmetry of the system, the sequential order of
coefficients is irrelevant. With the re-defined operator the effect
of decoherence reads
\begin{equation}
\mD(\varrho) = \varrho' + (1 - \Tr \varrho')\vrho,
\label{eq:inhomo-dop}
\end{equation}
where $\varrho=\ket{\psi}\bra{\psi}$ is the initial state, and the
non-vacuum part of the output state is
$\varrho'=\ket{\psi'}\bra{\psi'}$, with $\ket{\psi'}= D\ket{\psi}$.
Therefore, we can again reduce our problem to calculating the
evolution of unnormalized pure states, just as in the uniform case,
and use the non-unitary step operator $U''=DU'$ with the more
general noise operator.

Telling how well the algorithm performs under these conditions is a
complex task. First we give a lower bound on the probability of
measuring the target node, based on generic assumptions. To begin, we
separate the noise operator into two parts
\begin{equation}
D = \eta + D',
\label{eq:D-sep}
\end{equation}
where, for the moment, we leave $0\le\eta\le1$ undefined. As a
consequence of Eq.~(\ref{eq:D-inhomo}) the diagonal elements of $D'$
are $[D']_{dd}=\delta_d=\eta_d-\eta$, and the off-diagonal elements
are zero. From Eq.~(\ref{eq:inhomo-dop}) it follows that starting from
a pure state $\ket{\psi_0}$, after $t$ non-ideal steps the state of
the system can be characterized by the unnormalized vector
$\ket{\psi'(t)}$, which is related to the state obtained from the same
initial state by $t$ ideal steps as
\begin{equation}
\ket{\psi'(t)} = \eta^t\ket{\psi(t)} + \ket{r}.
\end{equation}
The expression of the residual vector $\ket{r}$ reads
\begin{equation}
\ket{r} = \sum_{k=1}^t (DU')^{t-k} D' \eta^{k-1} \ket{\psi(k)}.
\end{equation}
To obtain the probability of measuring the target state $\ket{x=0}$ we
have to evaluate the formula
\begin{equation}
p_n(\{\eta\},t) = \sum_{d=0}^{n-1} \left| \eta^t \braket{d,0}{\psi(t)} +
\braket{d,0}{r} \right|^2.
\label{eq:inhomo-prob-def}
\end{equation}
Due to the symmetry of the graph and the coins, we use e.g.\
Eq.~(\ref{eq:prob_uniform_approx}) and obtain $\braket{d,0}{\psi(t)}
\approx -\sin(\omega'_0t)/\sqrt{2n}$. To obtain a lower bound on
$p_n(\{\eta\},t)$ we note that the sum is minimal if
$\braket{d,0}{r}=\textrm{const}=K$ for every $d$ (we consider a worst
case scenario when all $\braket{d,0}{r}$ are negative). Now we assume
that the second term is a correction with an absolute value smaller
than that of the first term. For the upper bound on $K$, we use the
inequality
\begin{equation}
\sum_{d=0}^{n-1} \left| \braket{d,0}{r} \right|^2 \le \braket{r}{r}.
\end{equation}
The norm of $\ket{r}$ can be bound using the
eigenvalues of $U$, $D$, and $D'$. Let $\etamax = \max \Set{\eta_d |
d=0,\ldots,n-1}$ and $\delta_{\mathrm{max}} = \max
\Set{\left|\delta_d\right| | d=0,\ldots,n-1}$. Then we have
\begin{equation}
\braket{r}{r} \le \sum_{k=1}^t \etamax^{t-k}
\delta_{\mathrm{max}} \eta^{k-1} = \frac{\etamax}{\eta}
\frac{\delta_{\mathrm{max}}}{\etamax-\eta} (\etamax^t - \eta^t).
\end{equation}
Since $U$ is unitary, its contribution to the above formula is
trivial. Our upper bound on $|K|$ hence becomes $|K| \le 1/{\sqrt n}
(\etamax\delta_{\mathrm{max}}/\eta) (\etamax^t - \eta^t) / (\etamax -
\eta)$.  Combining the results, we obtain a lower bound on the
probability for measuring the target node,
\begin{multline}
p_n(\{\eta\},t) \ge \eta^{2t} \left\{ \sqrt{p_n^{(i)}(t)} \right. \\
  \left. - \frac{\etamax}{\eta} \frac{\delta_{\mathrm{max}}}{\etamax-\eta}
  \left[ \left(\frac{\etamax}{\eta}\right)^t -1 
  \right] \right\}^2,
\label{eq:p-lower-bound}
\end{multline}
where $p_n^{(i)}(t)$ stands for the corresponding probability of the
ideal (lossless) case. We maximize the lower bound with respect to the
arbitrary parameter $\eta$. The procedure can be carried out noting
that $\delta_{\mathrm{max}} = \max\{ \etamax-\eta,
\eta-\eta_{\mathrm{min}} \}$, thereby we find the maximum at
$\eta=\bar{\eta} \equiv (\etamax + \eta_{\mathrm{min}})/2$, yielding
the formula
\begin{equation}
p_n(\{\eta\},t) \ge \bar{\eta}^{2t}\left\{ \sqrt{p_n^{(i)}(t)} -
   ({\etamax}/{\bar{\eta}}) \left[
   \left({\etamax}/{\bar{\eta}}\right)^t -1 \right] \right\}^2.
\label{eq:opt-lower-bound}
\end{equation}
To interpret the formula (\ref{eq:opt-lower-bound}), we consider the
two terms in the curly braces separately. The first term returns the
success probability for uniform losses with transmission coefficient
$\bar\eta$. The second term may be considered as a correction term
that depends not only on some average value of the loss distribution,
but also on its degree of non-uniformity in a way that is reminiscent
of a mean square deviation. We observe that
Eq.~(\ref{eq:p-lower-bound}) provides a useful lower bound only for
$\{\eta\}$ distributions violating uniformity to only a small degree.
When the expression inside the curly braces becomes negative, the
assumption made on the magnitude of the second term of
Eq.~(\ref{eq:inhomo-prob-def}) becomes invalid, and therefore the
formula does not give a correct lower bound.

The estimated lower bound (\ref{eq:p-lower-bound}) decreases with
increasing degree of non-uniformity, in accordance with a naive
expectation. However, as we shall show later, numerical simulations
taking into account the full complexity of the problem provide
evidence to the contrary: departure from uniformity can result in
improved efficiency.

\begin{figure}

\includegraphics[scale=.67]{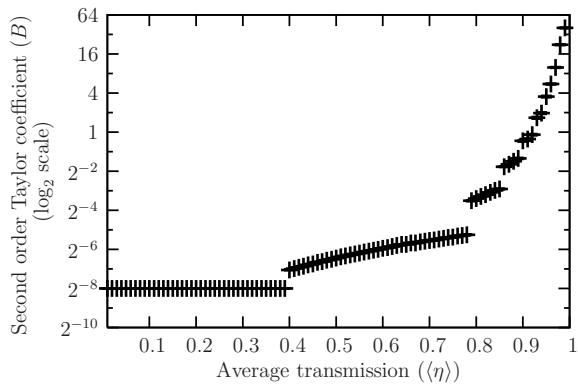}
\caption{The coefficients for the $\left<\eta\right>$ dependent second
  order term in the Taylor series expansion of
  $p_n^{\mathrm{\max}}(\{\eta\})$. The coefficients have been obtained
  by second order fitting to numerically generated values for a
  hypercube sized $n=8$. The confidence of each fit is represented on
  the graph by an error bar. It is clearly visible that the
  coefficients are always greater than $2^{-n}$. Another feature, that
  is more suggestive on a linear scale, is that the points between two
  steps seem to align into straight lines, with their slopes
  increasing with $\left<\eta\right>$.}
\label{fig:fit-taylor}
\end{figure}

Inspired by the appearance of the average loss rate in the lower bound
(\ref{eq:opt-lower-bound}), we introduce the mean and the variance of
the direction dependent losses,
\begin{equation}
\left<\eta\right> = \frac1n \sum_{d=0}^{n-1} \eta_d,
\quad
\mbox{and} 
\quad
Q = \frac1n \sum_{d=0}^{n-1} \delta_d^2.
\end{equation}
By using the Taylor expansion of the success probability function
$p_n^{\mathrm{max}}$ around the point $\eta_d=\left<\eta\right>$, the
deviations from the uniform loss case can be well estimated at small
degrees of non uniformity. Using the permutation symmetry of
$p_n^{\mathrm{max}}$ we can express the Taylor series as
\begin{equation}
p_n^{\mathrm{max}}(\{\eta\}) = p_n^{\mathrm{max}}(\left<\eta\right>) +
B Q^2 + C W^3 + O(\delta_d^4),
\label{eq:pmax-Taylor}
\end{equation}
where $W^3 = 1/n \sum_k \delta_k^3$. We notice that $Q$ may be
regarded as the mean deviation of $\{\eta\}$ as a distribution, and
hence it is a well-defined statistical property of the random
noise. In other words, as long as a second order Taylor expansion
gives an acceptable approximation, the probability of success depends
only on the statistical average and variance ($\left<\eta\right>$,
$Q$) of the noise and not on the specific values of $\{\eta\}$. Using
numerical simulations, we have determined the values of $B$ up to rank
$n=10$, and studied the impact of higher order terms.

The second order Taylor coefficients were determined by fitting over
the numerically obtained success probabilities at data points where
the higher order moments of the loss distributions were small. An
example plot of $B$ is provided on Fig.~\ref{fig:fit-taylor}, for a
system $n=8$. The higher order effects were suppressed by selecting
the lowest values of $W$ from several repeatedly generated random
distributions $\{\eta\}$. A general feature exhibited by all studied
cases is that the second order coefficients satisfy the inequality
\begin{equation}
B\ge 2^{-n}.
\label{eq:lowerb-second}
\end{equation}
It is remarkable that this tight lower bound depends only on the size
of the system. The dependence of $B$ on $\left<\eta\right>$ is
monotonous with discontinuities. We found the number of
discontinuities to be proportional to the rank $n$. Our numerical
studies have shown that the value of $B$ before the first
discontinuity is always a constant, and equal to the empirical lower
bound (\ref{eq:lowerb-second}).

\begin{figure}

\includegraphics[scale=.67]{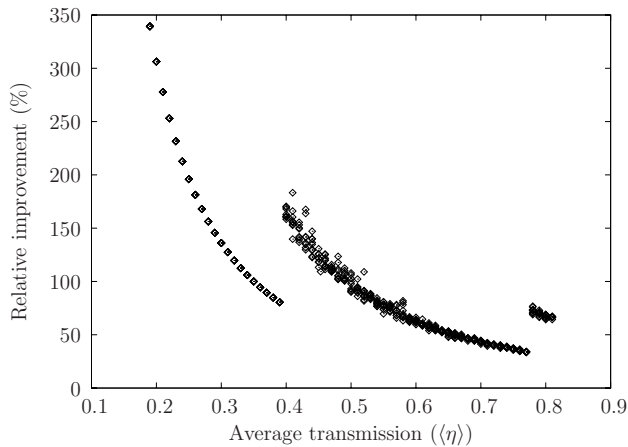}
\caption{The relative improvement of maximum success probability
  comparing direction dependent loss to uniform loss with identical
  average loss rates. The difference is measured as
  $[p_n^{\mathrm{max}}(\{\eta\}) -
  p_n^{\mathrm{max}}(\left<\eta\right>)] /
  p_n^{\mathrm{max}}(\left<\eta\right>)$. All points were generated at
  $Q=0.35$ and the entire available domain for
  $\left<\eta\right>$. The values of higher moments of $\{\eta\}$ are
  not restricted, therefore we see multiple points for certain
  $(\left<\eta\right>, Q)$ pairs. (Rank of the hypercube $n=9$.)}
\label{fig:max-prob-Q35}
\end{figure}

To plot the success probabilities corresponding to arbitrary random
coefficients we used the pair of variables $\left<\eta\right>$ and
$Q$. On these plots, the higher order terms cause a ``spread'' of the
appearing curves. A sample plot is displayed on
Fig.~\ref{fig:max-prob-Q35} where the relative improvement is compared
to the uniform case, in percentages. We observe a general increase of
efficiency as compared to the uniform case with the same average loss
rate. A general tendency is that for smaller values of
$\left<\eta\right>$ the improvement is larger, interrupted, however,
by discontinuities. These discontinuities closely follow those of the
second order coefficient $B$.

\begin{figure}

\includegraphics[scale=.67]{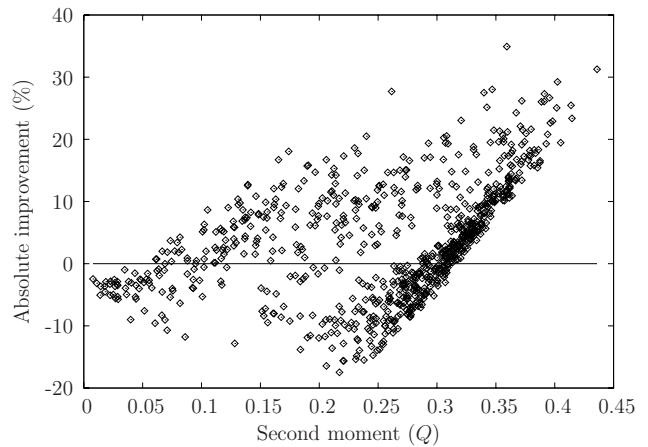}
\caption{The difference of the maximum success probabilities, in the
  presence of direction dependent loss with coefficients $\{\eta\}$,
  and in the presence of uniform loss with coefficient
  $\eta_{\mathrm{max}}=\max\{\eta\}$.  Physically, the non-uniform
  case can be obtained from the uniform case by introducing proper
  attenuation. We use the second moment $Q$ as a measure of deviation
  from the non-attenuated case. The vertical axis shows
  $[p_n^{\mathrm{max}}(\{\eta\}) -
  p_n^{\mathrm{max}}(\eta_{\mathrm{max}})] / p_n^{\mathrm{max}}
  (\eta_{\mathrm{max}})$ as a percentage. We observe a systematic
  improvement for higher values of $Q$. The plot corresponds to rank
  $n=7$ and $\eta_{\mathrm{max}}= 0.996\pm 0.001$. }
\label{fig:max-prob-max}
\end{figure}

The numerical studies, involving the generation of 1000 sets of
uniformly randomly generated transmission coefficients for each of the
systems of up to sizes $n=10$, indicate that with the help of
Eq.~(\ref{eq:lowerb-second}) the first two terms of the expansion
Eq.~(\ref{eq:pmax-Taylor}) can be used to obtain a general lower
bound:
\begin{equation}
p_n^{\textrm{max}}(\{\eta\}) \ge p_n^{\mathrm{max}}
(\left<\eta\right>) + 2^{-n}Q^2.
\end{equation}
The inequality implies that the overall contribution from higher order
terms is positive, or always balanced by the increase of $B$. The
appeal of this lower bound is that it depends only on the size of the
system $N=2^n$, and the elementary statistical properties of the noise
($\left<\eta\right>$, $Q$). Therefore, together with the formula
(\ref{eq:maxprob-approx}) for uniform loss, a straight-forward
estimation of success probability is possible before carrying out an
experiment.

Up to now, we concentrated on comparing the performance of the search
algorithm suffering non-uniform losses with those suffering uniform
loss with coefficient equal to the average of the non-uniform
distribution. Another physically interesting question is how
attenuation alone affects search efficiency. We can formulate this
question using the notations above as follows. Consider a randomly
generated distribution $\{\eta\}$ and compare the corresponding
success probability with the one generated by a uniform distribution
with transmission coefficient $\eta_{\mathrm{max}} = \max
\{\eta\}$. We chose $Q$ as a measure of how much an
$\eta_{\mathrm{max}}$ uniform distribution needs to be altered to
obtain $\{\eta\}$, and made the comparisons using the same set of
samples. A typical plot is presented on
Fig.~\ref{fig:max-prob-max}. It appears that as we start deviating
from the original uniform distribution, an initial drop of efficiency
is followed by a region where improvement shows some systematic
increase. However, it is still an open question, whether it is really
a general feature that for some values of $Q$ the efficiency is always
increased. On the other hand these plots provide clear evidence that
for a significant number of cases the difference
$p_n^{\mathrm{max}}(\{\eta\}) -
p_n^{\mathrm{max}}(\eta_{\mathrm{max}})$ is positive. In other words,
rather counter-intuitively, we can observe examples where increased
losses result in the improvement of search efficiency. Since the time
evolution with losses is non-unitary, the improvement cannot be
trivially attributed to the fact that the Grover operator is not the
optimal choice for the marked coin.

\section{Phase errors}
\label{sec:phase}

In the present section we discuss another type of errors typically
arising in optical multiport networks. These errors are due to
stochastic changes of the optical path lengths relative to what is
designated, and manifest as undesired random phase shifts. Depending
on how rapidly the phases change, we may work in two complementary
regimes. In the ``phase fluctuation'' regime the phases at each
iteration are different.  These errors can typically be caused by
thermal noise. In the ``static phase errors'' regime, the undesired
phases have slow drift such that on the time scale of an entire run of
the quantum algorithm their change is insignificant. The origin of
such errors can be optical element imperfections, optical
misalignments, or a slow stochastic drift in one of the experimental
parameters. Phase errors in the fluctuation regime have been studied
in Ref.~\cite{Kosik2006Quantum-walks-w} for walks on $N$ dimensional
lattices employing the generalized Grover or Fourier coin. The impact
of a different type of static error on the SKW algorithm has been
analyzed in Ref.~\cite{Li2006Gate-imperfecti}.

To begin the formal treatment, let $F$ denote the operator introducing
the phase shifts, and write it as
\begin{equation}
F(\{\varphi\}) = \sum_{d,x} e^{i\varphi_{dx}} \ket{d,x}\!\bra{d,x}.
\end{equation}
This operator is unitary, hence the step operator
\begin{equation}
 U(\{\varphi\}) = S F(\{\varphi\}) C',
\label{eq:phase-noise-U}
\end{equation}
that depends on the phases $\{ \varphi_{dx} \vert d=0..n-1,
x=0..2^n-1\}$ is unitary as well. In case of phase fluctuations, at
each iteration $t$ we have the parameters $\varphi_{dx}^{(t)}$ such
that all $\varphi_{dx}^{(t)}$ are independent random variables for
every $d$, $x$ and $t$, according to some probability distribution. In
case of static phase errors, $\varphi_{dx}^{(t)}$ and
$\varphi_{dx}^{(t')}$ are considered to be the same random variables
for every pair of $d$ and $x$.

The formalism of Ref.~\cite{Kosik2006Quantum-walks-w} can be applied
to the walk on the hypercube, and extended to the case of non-uniform
coins and position dependent phases. Namely, using the shorthand
notations $D=\left\{0,1,2,\ldots,n-1\right\}$ and
\begin{equation}
E(k,l) = \bigoplus_{j=l}^k e_{a_j},
\end{equation}
the state after $t$ iterations can be expressed as
\begin{multline}
\ket{\psi(\{\varphi\},t)} = \frac1{\sqrt{n2^n}} \sum_{x_0\in V}
(-1)^{\delta_{\xtg x_0}} \sum_{\underline{a}\in D^t}
  e^{i\varphi(\underline{a}, x_0)} \\
  \times \tilde \Xi_{\xtg} (\underline{a}, x_0) \ket{a_1, x_0 \oplus
    E(t,1)}, 
\end{multline}
where
\begin{multline}
\tilde \Xi_{\xtg} (\underline{a}, x_0) = \\
\prod_{j=1}^{t-1} \left(
  C^{(0)}_{a_ja_{j+1}} + [C^{(1)} - C^{(0)}]_{a_ja_{j+1}}
  \delta_{\xtg \oplus  x_0, E(t,j+1)}  \right),
\end{multline}
and $\varphi(\underline{a},x_0)= \sum_{j=1}^t
\varphi^{(t+1-j)}_{a_j,x_0 \oplus E(t,j-1)}$. For the standard SKW
algorithm, the coin matrices are $C^{(0)}_{aa'} = 2/n-\delta_{aa'}$
and $C^{(1)}_{aa'}=-\delta_{aa'}$, however, the SKW algorithm is
reported to work with more general choices of operators $C_{0/1}$
\cite{shenvi:052307}.

For the following study, we express the probability of finding
the walker at position $x$ after $t$ iterations as the sum
$p_n(x,\{\varphi\},t)=p_n^I(x,\{\varphi\},t) +
p_n^C(x,\{\varphi\},t)$, such that the incoherent and coherent
contributions are
\begin{eqnarray}
p_n^I(x,\{\varphi\},t) &=& \frac1{n2^n} \sum_{\underline{a}\in D^t} \left|
\tilde\Xi_{\txtg}(\underline{a}) \right|^2,
\label{eq:px-incoh} \\
p_n^C(x,\{\varphi\},t) &=& \frac1{n2^n} \sum_{\underline{a}\neq
  \underline{a}' } \Phi_{\underline{a}'}^* \Phi_{\underline{a}}
  \tilde\Xi_{\txtg}(\underline{a}')^* 
  \tilde\Xi_{\txtg}(\underline{a}) \delta_{a_1'a_1}, 
\label{eq:px-coh}
\end{eqnarray}
where $\txtg=\xtg \oplus x$. The appearing phase factors are
\begin{equation}
\Phi_{\underline{a}} = (-1)^{\delta_{\txtg, E(t,1)}}
e^{i\varphi(\underline{a}, x \oplus E(t,1))},
\end{equation}
and $\tilde \Xi_{\txtg}(\underline{a})=\tilde
\Xi_{\txtg}(\underline{a}, E(t,1))$, i.e. 
\begin{equation}
\tilde \Xi_{\txtg} (\underline{a}) =
\prod_{j=1}^{t-1} \left(
  C^{(0)}_{a_ja_{j+1}} + [C^{(1)} - C^{(0)}]_{a_ja_{j+1}}
  \delta_{\txtg, E(j,1)}  \right). 
\label{eq:Xi_tg}
\end{equation}
Note, that when the probability of finding the walker at the target
node $\xtg$ is to be calculated we must set $x=\xtg$, therefore, we
have $\txtg=0$.

In the following we shall show that the incoherent contribution is
constant,
\begin{equation}
p_n^I(x, \{\varphi\}, t) = \frac1{2^n},
\label{eq:px-coh-const}
\end{equation}
for any two unitary coins $C_{0/1}$. Consequently, $p_n^I$ is constant
also for balanced coins such as those in
Eq.~(\ref{eq:std-coin-pair}). The summations in
Eq.~(\ref{eq:px-incoh}) can be rearranged in increasing order of
indices of $a_j$, yielding
\begin{multline}
p_n^I(x, \{\varphi\}, t) = \\
  \frac1{n2^n} \sum_{a_1,a_2=0}^{n-1} \left|
  C^{(0)}_{a_1a_2} + [C^{(1)}- C^{(0)}]_{a_1a_2}
  \delta_{\txtg, e_{a_1}} \right|^2 \times \cdots  \\
  \times \sum_{a_{t-1}=0}^{n-1} \left| C^{(0)}_{a_{t-2}a_{t-1}} +
  [C^{(1)} - C^{(0)}]_{a_{t-2}a_{t-1}} \delta_{\txtg,
  E(t-2,1)} \right|^2 \\
  \times \sum_{a_t=0}^{n-1} \left| C^{(0)}_{a_{t-1}a_t} +
  [C^{(1)} - C^{(0)}]_{a_{t-1}a_t} \delta_{\txtg,
  E(t-1,1)} \right|^2.
\end{multline}
Since $E(t-1,1)$ depends on $a_j$ only when $j\leq t-1$, and due to
the unitarity of the coins $\langle a_{t-1}| C_0^{\dag} C_0 | a_{t-1}
\rangle = \langle a_{t-1}|C_1^{\dag} C_1 | a_{t-1} \rangle = 1$, the
summation over $a_t$ can be evaluated and we obtain $1$. Hence, we see
that $p_n^I(x, \{\varphi\}, t) = p_n^I(x, \{\varphi\}, t-1)$, and this
implies Eq.~(\ref{eq:px-coh-const}) by induction. 

\begin{figure}

\includegraphics[scale=.67]{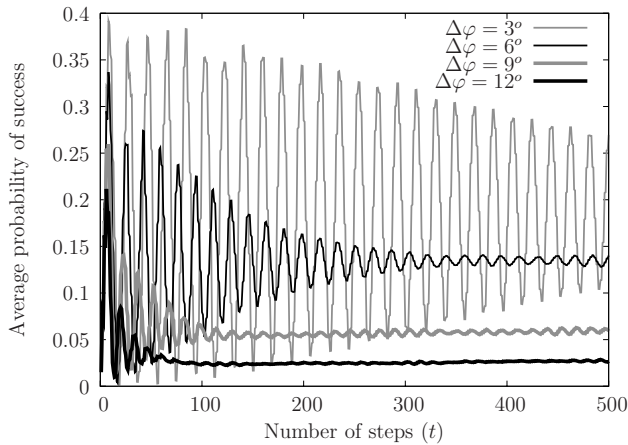}
\caption{The averaged (1000 samples) probability of measuring the
  target node, when $n=6$ and $\Delta\varphi = 3^o, 6^o, 9^o,
  12^o$. The tendency of the success probability to a constant,
  non-zero value can be observed on this numerically obtained plot. It
  is also observable that a larger variance results in a smaller
  asymptotic value.}
\label{fig:phase-noise-evolution}
\end{figure}

The average probability $\bar{p}_n(x,t)$ of finding the walker at node
$x$ is obtained by averaging the random phases according to their
appropriate probability distribution. Using
Eq.~(\ref{eq:px-coh-const}) this probability can be expressed as
\begin{equation}
\bar{p}_n(x,t) = \left<p_n(x, \{\varphi\}, t) \right> =
\frac1{2^n} + \left< p_n^C(x, \{\varphi\}, t) \right>,
\end{equation}
where $\left<\ldots\right>$ denotes taking the average for each random
variable $\varphi^{(t)}_{dx}$ in case of phase fluctuations, and for
each $\varphi_{dx}$ in case of static phase errors. It is reasonable
to assume that each random variable has the same probability
distribution. To analyze the impact of phase errors on the search
efficiency, we study the behaviour of the coherent term $\left<
p_n^C(x, \{\varphi\}, t) \right>$ for different random distributions.

In case of phase fluctuations characterized by a uniform distribution,
the coherent term immediately vanishes and we obtain $\bar{p}_n(x,t) =
1/2^n$.  This case can be considered as the classical limit of the
quantum walk. Therefore, we conclude that the classical limit of the
SKW algorithm is not a search algorithm, independently of the two
unitary coins used. 

Assuming a Gaussian distribution of random phases is motivated by the
relation of each phase variable $\varphi$ to the optical path
length. The changes in the optical path lengths which introduce phase
shifts are not restricted to a $2\pi$ interval. In what follows, we
assume that the random phases have a zero centered Gaussian
distribution with a variance $\Delta\varphi$.

We arrive at the classical limit even when the phase fluctuations have
a finite width Gaussian distribution, simply by repeatedly applying
the time evolution operator $U\{\varphi\}$. For such Gaussian
distribution, the coherent term exhibits exponential decrease with
time, a behaviour also confirmed by our numerical calculations.

In the static phase error regime the mechanism of cancellation of
phases is different than in the fluctuation regime, and more difficult
to study analytically. For uniform random distribution we expect a
sub-exponential decay of the coherent term to zero. For a zero
centered Gaussian distribution with variance $\Delta \varphi$ we
performed numerical simulations using the standard two coins of
Eq.~(\ref{eq:std-coin-pair}).

\begin{figure}

\includegraphics[scale=.67]{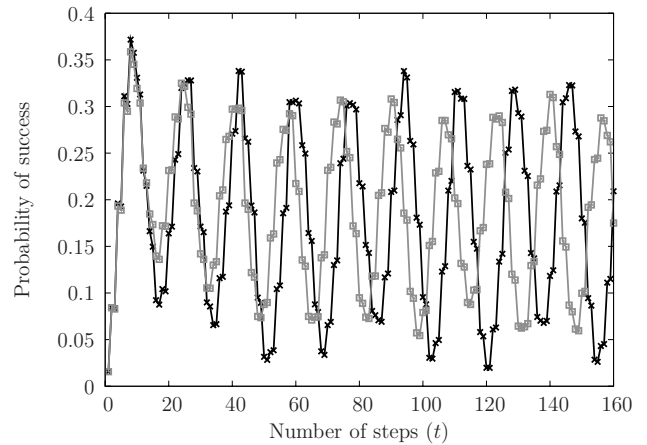}
\caption{Time dependence of success probabilities for two different
  phase configurations, numerically calculated for a system of rank
  $n=6$. The difference in frequencies of the major oscillations is
  clearly observable for larger times.}
\label{fig:phase-noise-diff}
\end{figure}

The numerical results for the success probability $\bar{p}_n(\xtg, t)$
for several values of $\Delta\varphi$ are plotted on
Fig.~\ref{fig:phase-noise-evolution}. The data points were obtained by
calculating success probabilities for $1000$ randomly generated phase
configurations and taking their averages at each time step $t$.

\begin{figure}

\includegraphics[scale=.67]{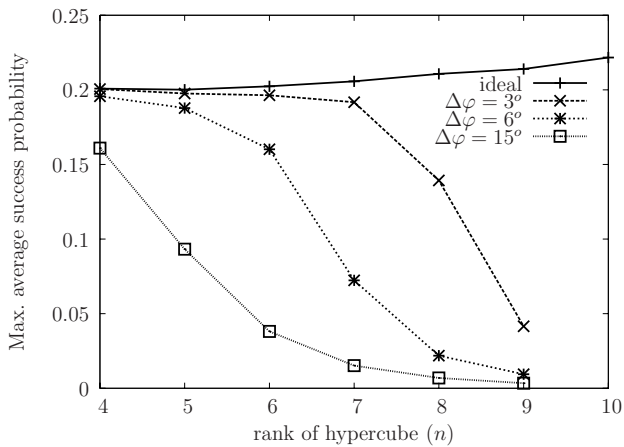}
\caption{The long time stationary values of the success probability
   (obtained by averaging over 1000 samples) against the size of the
   search space, for $\Delta\varphi=0^o,3^o,6^o,15^o$.}
\label{fig:phase-noise}
\end{figure}

By studying the repetition of the random phase configuration we come
to several remarkable conclusions. First, the time evolution of the
success probability tends (on a long time scale, $t\gg t_{f}$) to a
finite, non-zero constant value. Consequently, being subject to static
phase errors, the SKW algorithm retains its utility as search
algorithm. Second, the early steps of the time evolution are
characterized by damped oscillations reminding of a collapse. Third,
the smaller the phase noise the larger is the long time stationary
value to which the system evolves. We have plotted the stationary
values obtained by numerical calculations, against the rank of the
hypercube on Fig.~\ref{fig:phase-noise}.

Better insight into the above features can be gained by examining the
shape of the individual runs of the algorithm with the given random
phase configurations. As it can be seen on
Fig.~\ref{fig:phase-noise-diff}, the success probabilities for
different runs display the typical oscillations around a non-zero
value. They differ slightly in their frequencies depending on the
random phases chosen, hence when these oscillations are summed up we
get the typical collapse behaviour. Also, since these frequencies
continuously fill up a band specified by the width of the Gaussian, we
expect no revivals to happen later. For higher order hypercubes the
success probability drops almost to zero for already very moderate
phase errors, resembling a behaviour seen on
Fig.~\ref{fig:plot_prob_dimless}.

\section{Conclusions}
\label{sec:conclusions}

We studied the SQRW implementation of the SKW search algorithm and
analyzed the influence on its performance the two most common type of
disturbances, namely photon losses and phase errors. Our main result
for the photon loss affected SQRW search algorithm is that the
introduction of non-uniform distribution of the loss can significantly
improve the search efficiency compared to uniform loss with the same
average. In many cases, even the sole increase of losses in certain
directions may improve the search efficiency. Mostly based on
numerical evidence we have set a lower bound for the search
probability as a function of the average and variance of the randomly
distributed direction dependent loss.

We concentrated our analysis on two complementary regimes of phase
errors. When the system is subject to rapid phase fluctuations, the
classical limit of the quantum walk is approached. We have shown that
in this limit the SKW algorithm loses its applicability to the search
problem for any pair of unitary coins. On the other hand, we showed
that when the phases are kept constant during each run of the search,
the success rate does not drop to zero, but approaches a finite
value. The effect in its mechanism is reminiscent to exponential
localization found in optical networks \cite{TJS}. Therefore, in the
long-time limit, static phase errors are less destructive than rapidly
fluctuating phase errors.

\acknowledgments{ Support by the Czech and Hungarian Ministries of
  Education (CZ-2/2005), by MSMT LC 06002 and MSM 6840770039 and by
  the Hungarian Scientific Research Fund (T049234 and T068736) is
  acknowledged.  }

\bibliographystyle{apsrev}

\end{document}